\documentclass{article}
\usepackage{spconf,amsmath,graphicx}

\usepackage{enumitem}
\setlist{nosep, leftmargin=14pt}

\usepackage{mwe} 
\usepackage{booktabs}
\usepackage{hyperref}

\title{Exploiting segmentation labels and representation learning to forecast therapy response of PDAC patients}
\name{Alexander Ziller$^{1,*}$\thanks{$^{*}$equal contribution}, Ayhan Can Erdur$^{1,*}$, Friederike Jungmann$^{1}$}{Daniel Rueckert$^{1,2}$, Rickmer Braren$^{1,4}$, Georgios Kaissis$^{1,3}$}

\address{$^{1}$ Klinikum Rechts der Isar, Technical University of Munich\\
    $^{2}$ Imperial College London\\
    $^{3}$ Helmholtz Zentrum M\"unchen\\
    $^{4}$ German Cancer Consortium DKTK, Partner Site Munich}
\begin{document}
%
\maketitle
\begin{abstract}
The prediction of pancreatic ductal adenocarcinoma therapy response is a clinically challenging and important task in this high-mortality tumour entity. The training of neural networks able to tackle this challenge is impeded by a lack of large datasets and the difficult anatomical localisation of the pancreas. Here, we propose a hybrid deep neural network pipeline to predict tumour response to initial chemotherapy which is based on the \textit{Response Evaluation Criteria in Solid Tumors} (RECIST) score, a standardised method for cancer response evaluation by clinicians as well as tumour markers, and clinical evaluation of the patients. We leverage a combination of representation transfer from segmentation to classification, as well as localisation and representation learning. Our approach yields a remarkably data-efficient method able to predict treatment response with a ROC-AUC of $63.7\%$ using only $477$ datasets in total. 
\end{abstract}

\section{Introduction}
Pancreatic ductal adenocarcinoma (PDAC) is among the tumour entities with the highest mortalities worldwide. Its diagnosis is challenging and relies fundamentally upon high-resolution imaging such as computed tomography (CT) imaging. 
Imaging likely yields a wealth of information about the tumour no imaging facts beyond the localisation and size of the lesion, its anatomical surroundings as well as the presence or absence of tumour spreading. However, this information is not considered for diagnosing PDAC and evaluating its treatment response according to the current guidelines \cite{pdac_guidelines}.
With the rise of deep learning, models aiming to extract the aforementioned information and assist physicians in the assessment of such tumours have begun to appear \cite{ardila2019end}. For example, machine learning-assisted prediction of molecular tumour subtype and patient prognosis under therapy have recently been presented \cite{jungmann2021prediction,kaissis2019machine,kaissis2019machine2}.
As many PDAC are discovered at later stages where a primary resection is infeasible, chemotherapy is a first-line treatment for many patients, either to shrink the tumour prior to an operation or as a palliative indication. In this setting, the prediction of tumour response to initial treatment is a particularly interesting end-point.   
For example, deep neural networks trained to predict treatment response conditioned on the specific choice of chemotherapy can in the future be employed to select the personalised treatment with the highest success probability and/or best outcome. 
In clinical practice, standardised metrics of tumour response already exist: The \textit{Response Evaluation Criteria in Solid Tumors} (RECIST) \cite{schwartz2016recist,litiere2017recist} are a framework to classify whether a tumour is progressive, stable or regressive under therapy.
Although RECIST can be used to derive labels for training neural networks, a number of challenges have so far inhibited the success of such machine learning approaches: 
\begin{itemize}
    \item As in most medical problems, sufficiently large datasets to train deep neural networks are difficult to acquire;
    \item Abdominal CT scans cover a large volume of the body, whereas the pancreas (and potential extrapancreatic tumour manifestations) represent(-s) a proportionally small area of interest. Thus, there exists a risk that models which operate on the entire CT volume miss relevant information in the scan;
    \item Compounding the two problems, methods which are capable of focusing on relevant areas within the image (e.g. attention-based architectures) are --in practice-- even more data-hungry \cite{dosovitskiy2020vit}.
\end{itemize}
\textbf{Our approach}\newline
To overcome the problem of data scarcity, we propose to leverage knowledge from related public datasets \textit{but from an unrelated task}. Concretely, we propose a multi-step pipeline incorporating representations learned from a segmentation task into our classification model. We realise this through initially training a hybrid segmentation and object detection model to pre-crop the CT volume to the pancreatic region. Imitating the human approach to geometric pattern matching, we feed the segmentation masks into the classification model as assistive information. Moreover, our pipeline allows us to re-use the segmentation weights for the classification task, further increasing data efficiency. Finally, we use triplet loss-guided learning to enhance the quality of the classification model's intermediate representations. Our contributions can be summarised as follows:
\begin{itemize}
    \item We propose an integrated deep learning pipeline based on cascading a slice classifier and a segmentation model and complemented by transfer and representation learning. This results in --to our knowledge-- the first successful approach to response prediction to chemotherapy from baseline CT scans in PDAC using fewer than $500$ total datasets;
    \item We find that the incorporation of each of the intermediate steps outlined above in the pipeline yields substantial improvements in terms of performance and/or efficiency;
    \item We perform a detailed ablation study of the aforementioned intermediate components.
\end{itemize}
\textbf{Prior work}\newline
So far, very few works have tackled the challenging task of treatment response prediction in PDAC. For example, \cite{lu2021deep} break the task down to a comparison instead of a forecasting task. Other relevant previous approaches estimate diameters of lesions, which are an important part of the RECIST criteria \cite{tang2021lesion}. In terms of methodology, the closest work to our approach are \cite{azimi2022improving}, who use lung segmentation models to classify chest X-rays. Similarly to the second stage of our approach this work uses an upstream model to predict a segmentation, which is used to generate a cropping bounding box over a region of interest. Moreover, \cite{saleem2022deep, zhou2021multi} simultaneously learn segmentation and classification tasks. However, these works have both labels available for both tasks, whereas we learn the tasks disjointly with only one label being available per dataset and task. Finally, our approach to utilise autoencoder-like architectures for representation learning has also been explored in \cite{bengio2006greedy,kohlbrenner2017pre}, however --unlike us-- not with segmentation architectures.

\section{Methods}
\begin{figure}
    \centering
    \includegraphics[width=0.40\textwidth]{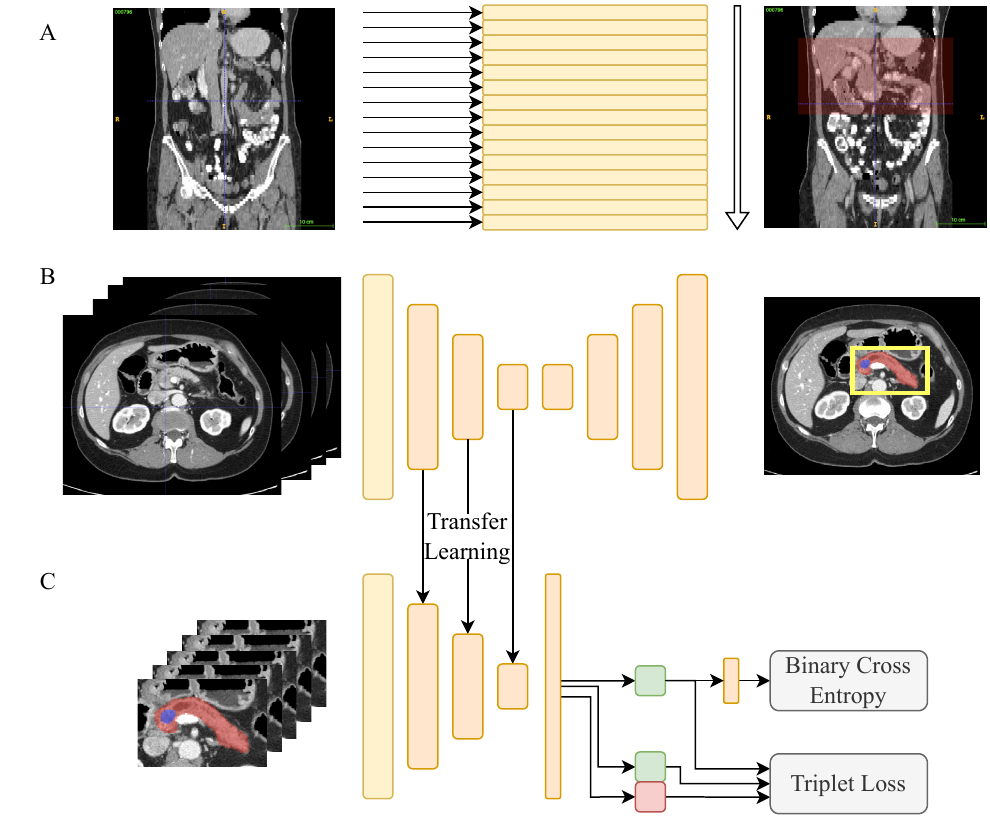}
    \caption{Overview of our approach. A: We classify each slice along the $z$-direction whether it contains pancreatic tissue in a $2.5$-D approach. B: We use a segmentation network, to crop in $x$ and $y$-direction, as well as feeding segmentation masks to the classification model. We re-use the network weights of the feature extractor part for the classification stage. C: We train a classifier, first using a triplet loss and afterwards using binary cross entropy. Images are extracted from the Medical Segmentation Decathlon dataset.}
    \label{fig:pipeline}
\end{figure}
As discussed above, we propose a multi-stage pipeline consisting of three sequential convolutional neural networks (CNNs) to increase the chemotherapy response classification performance. Furthermore, we evaluate the effect of each individual component in our pipeline. 

\subsection{Dataset and Tasks}
Our approach is motivated by the fact that we only have access to two datasets with disjoint labels.
For the first two stages, we are dependent on segmentation labels. As these are not present in the classification dataset, we use the public \textit{Medical Segmentation Decathlon} \cite{antonelli2022medical} dataset. It contains $N=281$ abdominal CT scans alongside per-voxel labels whether a point is part of the pancreas, cancerous tissue or background. 

We use an in-house dataset of $N=477$ CT scans for the classification task, where we predict treatment response labels, grouped into a binary prediction of progressive disease vs. stable and regressive diseases. As the clinically important distinction is whether patients experienced tumour progression or not, we perform a binary prediction on progressive cases (PD) against all other stable and regressive cases. This yields a class distribution of $N=171$ progressive vs $N=306$ stable or regressive CT scans. We split the dataset into a training subset of $N=420$ ($151$PD/$269$ other) samples and a testing subset of $N=57$ ($20$PD/$37$ other) samples.

In the following, we describe the general model we train, as well as additional pre-processing by segmentation models and representation learning.

\subsection{Classification}
Our overall goal is to train a network that predicts the treatment response in form of a RECIST label classification. Acting as the baseline of our work, we train a voxel-based $3$-D ResNet50 CNN to predict the treatment response label of the scan. We then evaluate the effect of the subsequent components in our integrated model cascade on the classification performance, as discussed in Section \ref{sec:exp}.

\subsection{Stage I: Slice Classification}
In order to limit the search space along the $z$-Dimension of a CT scan, we train a binary classifier which predicts for each slice whether it contains at least one pancreas or tumour pixel. Compared to a full segmentation model, this has the advantage of being very robust, and the model can be trained very efficiently and evaluated on large datasets while re-using the segmentation labels. To account for the three-dimensional structure of the task, we apply a Long Short-Term Memory (LSTM) cell after a $2$-D encoder. The LSTM consecutively receives extracted image features and thus holds information about previous slices, turning the model into a $2.5$-D classifier. As the $2$-D encoder, we employ a ResNet50 pre-trained on ImageNet \cite{deng2009imagenet}. The single LSTM cell following the encoder has $1024$ as the hidden size. Our $2.5$-D approach is based on the notion that we can leverage prior knowledge of the classification target, i.e. that it is a single continuous object within the scan. Moreover, we post-process the predictions and automatically include slices between positively classified slices, as we know that these contain the organ of interest. 

\subsection{Stage II: Segmentation}
As the second pipeline component, we train a $3$-D segmentation network.For this, we use a voxel-based architecture, \textit{DynU-Net} from MonAI \cite{monai} model zoo. The segmentation step is more expensive in terms of training and inference time, as well as more prone to inaccuracies compared to our slice classification model. However, its contributes considerable utility to the overall pipeline. It allows us to reduce the search space not only in the $z$-direction as above, but also in the $x$ and $y$ directions by converting the segmentation to a bounding box around the organ. Moreover, we later feed the predicted segmentations as additional image channels to the next stage and thus introduce additional (assistive) information. This approach is motivated by the human geometric pattern-matching approach, where the detection of specific shapes facilitates overall recognition. Lastly, we can reuse the network weights of the encoder for the next stage of the pipeline and profit from task-specific pre-trained weights as a \textit{transfer learning} scheme. The re-utilisation of weights allows us to reduce the size of the RECIST classifier network and thus the computational power and training and inference time decrease.

\subsection{Representation learning}
As the final component of our pipeline, we evaluate the benefit of a triplet loss term \cite{schroff2015facenet}, such that the intermediate representations of samples of the same class are close, whereas the representations of different classes are further apart in the latent space. For this, we always sample an \textit{anchor} image, a \textit{positive} image from the same class and a \textit{negative} image from the negative class. We calculate the loss in terms of $L_2$-distance of the intermediate representations after the feature extractor stage of the model (i.e. immediately before the linear layers).
We use this loss term to condition the network on this task in a first training stage, and only in a second stage trained the classification part using a binary cross-entropy loss. 
\section{Experiments}\label{sec:exp}
\begin{table*}[h]
    \centering
    \begin{tabular}{llllllll}
        \toprule
        Method & \multicolumn{2}{c}{MCC} & \multicolumn{2}{c}{Accuracy} & \multicolumn{2}{c}{AUC-ROC} \\
         &  $\mu$ & $\sigma$& $\mu$ & $\sigma$& $\mu$ & $\sigma$ \\ \midrule
        Baseline &  $5.6\%$& $4.4\%$&$41.0\%$&$4.5\%$&$45.5\%$&$7.9\%$ \\
        + slice ($z$) cropping &$15.5\%$&$7.5\%$&$54.2\%$&$9.8\%$&$57.5\%$&$4.4\%$ \\
        + informed $x$/$y$ cropping &$18.8\%$&$4.0\%$&$61.1\%$&$4.4\%$&$59.2\%$&$3.1\%$ \\
        + segmentation forwarding &$21.5\%$&$1.4\%$&$53.3\%$&$4.9\%$&$59.3\%$&$3.5\%$\\
        + transfer learning &$21.6\%$&$4.7\%$&$64.1\%$&$3.1\%$&$59.8\%$&$3.6\%$ \\
        + triplet loss & $33.1\%$ & $2.2\%$&$67.2\%$&$3.6\%$&$63.7\%$&$2.6\%$ \\
        \bottomrule
    \end{tabular}
    \caption{Classification metrics on our validation data. Here, MCC stands for the \textit{Matthew's Correlation Coefficient}, \textit{accuracy} is the binary classification accuracy, \textit{AUC-ROC} is the area under the receiver-operator characteristic curve. $\mu$ denotes the mean result, $\sigma$ denotes the standard deviation of the results for $5$ runs with different random seeds.}
    \label{tab:results}
\end{table*}

\subsection{Slice Classification and Segmentation}
Both models are trained separately on the Medical Segmentation Decathlon \cite{antonelli2022medical} dataset. We set a maximum of $200$ epochs but monitor appropriate metrics and stop the training at convergence. On the held-out validation set, our slice classification network reaches \textbf{90.7\% accuracy} on predicting the presence of pancreatic tissue in a slice. The DynU-Net segmentation network succeeds to segment the pancreas with a \textbf{Dice score} of \textbf{77\%} and the tumour with \textbf{40\%}.
\subsection{Classification Results}
\begin{figure}[h]
    \centering
    \includegraphics[width=0.3\textwidth]{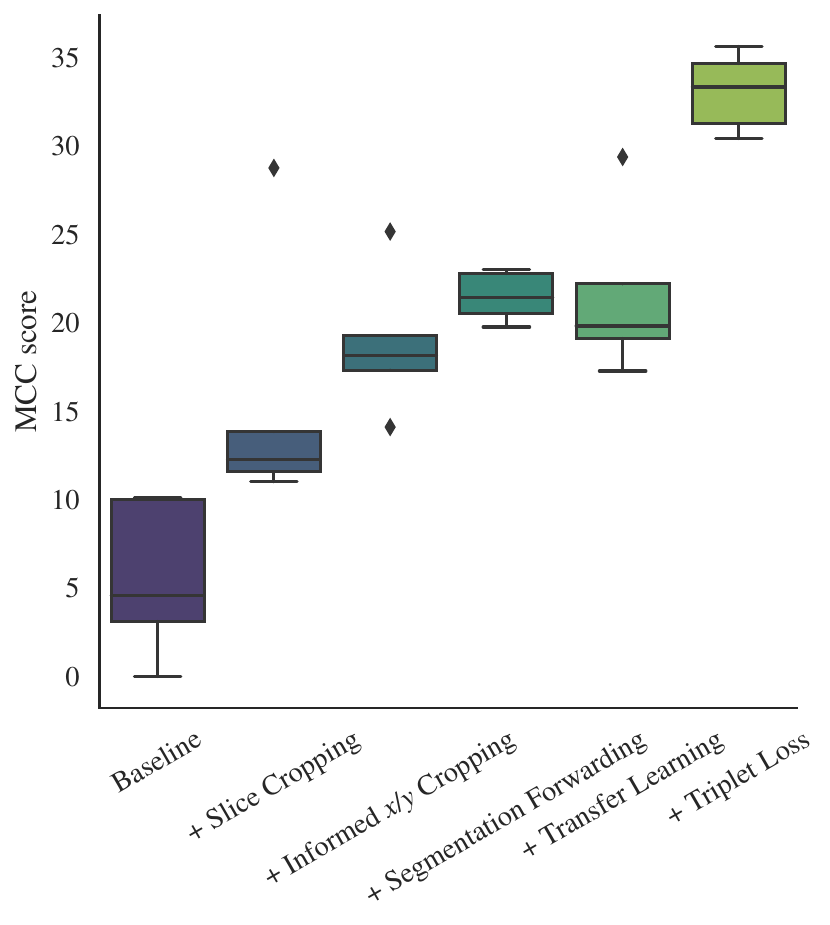}
    \caption{Box-and-whisker plots of the MCC scores for each integrated pipeline component averaged over $N=5$ runs.}
    \label{fig:violin}
\end{figure}

In the following, we report the classification performance of the individual pipeline components as well as the overall pipeline in our experimental evaluation. We use the Matthew's correlation coefficient (MCC) \cite{matthews1975comparison} as the main metric of our evaluation due to its clear advantages over other metrics \cite{chicco2020advantages}, most prominently, its robustness to class imbalance. For the sake of completeness, we also report the binary accuracy (unweighted) as well as the area under the receiver-operator characteristic curve (AUC-ROC), however we stress that these results should be interpreted with caution as they are susceptible to class imbalance. 
We perform each experiment with $5$ random seeds and report the standard deviation of the runs. As a baseline, we train our classifier model without modifications on the input CT-scan. We use an input shape of $256 \times 256 \times 256$ for this baseline. In the following configurations we can reduce this shape without losing the image resolution thanks to learned cropping. We compare this set of experiments to each incremental change introduced above. Cropping not only reduces the input to relevant features for the classification network, but also allows to lift the resolution limit. In the case where we only use a slice classifier to crop in the $z$-dimension we perform centre cropping in the $x$ and $y$ direction, whereby we exploit the fact that the pancreas is located centrally in the abdomen. 
Our findings are summarised in \autoref{tab:results}. The baseline model predicts almost random outputs, with an MCC score close to zero and a high standard deviation between the different runs. A simple pre-processing step to classify each slice whether it contains the pancreas and thus allowing a cropping operation of the scan along the $z$-direction already improves the results considerably (\textbf{slice cropping}). We suspect that an important reason to this is that by cropping, we can train on higher resolution input images without reaching hardware limits and reduce the scan to its relevant anatomical regions. However, the deviation between runs within this setting is still remarkably high. This variation is reduced by using a segmentation model to additionally crop in the $x$ and $y$ directions (\textbf{informed $x$/$y$-cropping}) instead of na\"ive centre cropping, and even further by adding the segmentations as input to the classifier (\textbf{segmentation forwarding}). 
Reusing the feature extractor of the segmentation \textit{DynU-Net} model and \textbf{transfer learning} it as a classification network did not yield a notable improvement in terms of test metrics. However, this model is much smaller and thus less computational intensive, which allowed a $2.5\times$ average training speed-up as well as a substantially decreased energy consumption. 
Lastly, the introduction of an additional \textbf{triplet loss} term on the intermediate representations of our classification network further improves the metrics. We reach a final MCC of $33.1\%$ on average with a standard deviation of $2.2\%$. MCC is a more pessimistic metric than accuracy, which is at $67.2\%$ ($36-41/57$ patients over 5 runs), but very robust to class imbalances, which is important in our dataset. These results are visually summarised in \autoref{fig:violin}. 

\section{Conclusion}
In this work we analyse the effects of iterative changes to the problem of treatment response classification of PDAC on baseline CT scans. We showed that exploiting available segmentation labels, as well as using representation learning can yield a large improvement in classification results. We hope that this work is a first step towards accurately predicting the chemotherapy response of PDAC patients and thus leading to an improved personalised treatment scheme. 

\section{Acknowledgments}
This work was supported by the European Research Council (Deep4MI - 884622), German Research Foundation, Priority Programme SPP2177, Radiomics: Next Generation of Biomedical Imaging, 
German Cancer Consortium Joint Funding UPGRADE Programme: Subtyping of Pancreatic Cancer based on radiographic and pathological Features. The funders played no role in the design or execution of the study, nor on the decision to prepare or submit the manuscript.
We used Jakob Richter's Punkreas as basis for parts of our code (\url{https://gitlab.com/sanddorn/punkreas}). 
\bibliographystyle{IEEEbib}
\bibliography{strings,refs}

\end{document}